\documentclass[aps,pra,twocolumn,showpacs,superscriptaddress]{revtex4}
\usepackage{graphicx}
\usepackage{amsmath}
\usepackage{amssymb}

\input{epsf}
\begin{document}

\title{Correlations of the upper
branch of 1d harmonically trapped 
two-component Fermi gases}

\author{Seyed Ebrahim Gharashi}
\affiliation{Department of Physics and Astronomy,
Washington State University,
  Pullman, Washington 99164-2814, USA}
\author{D. Blume}
\affiliation{Department of Physics and Astronomy,
Washington State University,
  Pullman, Washington 99164-2814, USA}
\affiliation{ITAMP, Harvard-Smithsonian Center for Astrophysics,
60 Garden Street, Cambridge, Massachusetts 02138, USA}

\date{\today}

\begin{abstract}
We present highly-accurate 
energy spectra and eigen functions of small 1d harmonically trapped
two-component Fermi gases with interspecies
$\delta$-function interactions, and analyze the correlations of
the so-called upper branch
(i.e., the branch that describes
a repulsive Fermi gas consisting of atoms but no
molecules) for positive and negative coupling
constants. 
Changes of the two-body correlations as a function of the
interspecies coupling strength reflect the competition of the
interspecies interaction and the effective repulsion
due to the Pauli exclusion principle, and are interpreted
as a few-body analog
of
a transition from a non-magnetic to a magnetic  phase.
Moreover,
we show that the eigenstate $\psi_{\rm{adia}}$
of the infinitely strongly-interacting 
system with $|n_1+n_2|>2$ and $|n_1-n_2| < n$ 
($n_1$ and $n_2$ denote the number
of fermions of components 1 and 2,
respectively), which is reached experimentally by adiabatically
changing the system parameters,
does not, as previously proposed, coincide with the wave function 
$\psi_{\rm{G}}$ obtained by applying a generalized
Fermi-Fermi mapping function to the 
eigen function
of the non-interacting  single-component Fermi gas.
\end{abstract}


\maketitle

1d systems serve as powerful models
whose study provides insights into fundamental phenomena such as 
gas dynamics, electron transport, 
Cooper pairing and
superconductivity~\cite{giamarchi,gogolin,RMP1,RMP2}.
In the special case where the interactions between the particles
are modeled by zero-range $\delta$-functions,
the quantum mechanical problem becomes integrable.
The integrability has many important consequences.
For example, 
1d systems with $\delta$-function interactions
can,
if external forces are absent and periodic boundary
conditions are imposed, 
be solved via the Bethe ansatz~\cite{betheansatz}. 
Another consequence of the integrability is the fact that
a single-component Bose gas with infinitely strong
$\delta$-function interactions behaves like an impenetrable
Bose gas, referred to as Tonks-Girardeau 
gas~\cite{girardeau65,olshanii,bloch,weiss}.
The corresponding bosonic wave function has similarities
with that of a gas of non-interacting (NI) fermions; in fact, the bosonic
wave function can be mapped to the fermionic wave function
via a Bose-Fermi mapping~\cite{olshanii,mapping05_2,granger,mapping05_1}. 
This Bose-Fermi duality has wide ranging applications.
In studies of lattice Hamiltonian, e.g., it implies that
bosonic creation and annihilation operators can be 
mapped to fermionic ones, and vice versa.

Given the success of the Bose-Fermi duality for
single-component Bose and Fermi gases, it is
intriguing to ask whether analogous dualities 
exist for
trapped multi-component gases with interspecies $\delta$-function
interactions with coupling strength $g$.
This question is not only of fundamental interest
but directly relevant to ongoing cold atom experiments
on effectively 1d two-component Fermi 
gases~\cite{zuernPRL,zuernThesis}.
Indeed, a generalized Fermi-Fermi mapping was recently
formulated for harmonically 
trapped two-component Fermi gases with infinitely large
interspecies
$\delta$-function interactions.
The generalized
Fermi-Fermi mapping~\cite{girardeauRapid} states that an eigenstate
$\psi_{\rm{G}}$ of the trapped two-component Fermi gas
with $|g| \rightarrow \infty$ can be obtained, for any $n$ ($n=n_1+n_2$), 
by applying a mapping function $M_{\rm{FF}}$ to the 
eigen function $\psi_{\rm{ideal}}$
of the NI harmonically
trapped one-component Fermi gas, i.e.,
$\psi_{\rm{G}}=
M_{\rm{FF}} \psi_{\rm{ideal}}$.
This Letter shows
that the states $\psi_{\rm{G}}$, 
constructed according to the generalized Fermi-Fermi mapping,
do in general not agree 
with the eigenstates $\psi_{\rm{adia}}$ of the two-component
Fermi gas,
which
emerge by adiabatically evolving the system Hamiltonian
from the NI to the infinitely strongly-interacting regime.
For $n>2$ and $|n_1-n_2| < n$, 
the eigenenergies 
for states with a given parity
for $|g| \rightarrow \infty $ are 
degenerate~\cite{guan},
thereby explaining how $\psi_{\rm{G}}$ can be an eigenstate but not 
coincide with any of the
states that are reached by performing an adiabatic sweep.

We
also calculate the pair correlation functions
of the upper branch of the
$(n_1,n_2)=(2,1)$, $(3,1)$ and $(2,2)$ systems.
The energy of the upper branch, which corresponds to a
metastable repulsive atomic gas, lies above that of the 
NI system, and is populated by starting from the
NI regime and turning on repulsive interspecies interactions.
The changes of the structural correlations
as
the
coupling constant is changed from small and positive, to
infinitely large, to small and negative
reflect the competition between the interspecies
interactions and the Pauli pressure introduced by
the anti-symmetry requirement of the wave function under
the exchange of identical fermions.
The expectation value of the intraspecies distance coordinate
exhibits a maximum at $g_{\rm{c}}$.
For $-1/g \lesssim -1/g_{\rm{c}}$, the interactions are 
``weaker'' than the Pauli exclusion principle and the
expectation value of the intra- and 
interspecies distances increase with increasing $-1/g$.
For $-1/g \gtrsim -1/g_{\rm{c}}$, in contrast, the interactions become so strong that
the system prefers to reduce the distance between like particles
with increasing $-1/g$.
These structural changes are interpreted as constituting
a smooth few-body
analog of the transition from a non-magnetic to
a magnetic phase.
The question whether
3d atomic 
two-component Fermi gases undergo,
if ``driven up'' the upper branch, a transition from
a paramagnetic to an itinerant ferromagnetic 
phase, as described by the Stoner model~\cite{stoner},
has recently been studied 
extensively experimentally and 
theoretically for 3d
two-component Fermi 
gases~\cite{ketterle,ketterle2,zwierlein,ho,demler,liu10,pilati,chang}. 

We consider $n$ 1d fermions 
with mass $m$ and position coordinates $z_j$.
Assuming 
interspecies
$\delta$-function interactions with coupling strength
$g$, the Hamiltonian reads
\begin{eqnarray}
H = \sum_{j=1}^n 
\left( \frac{-\hbar^2}{2m} \frac{\partial^2}{\partial z_j^2}
+ \frac{1}{2}m \omega^2 z_j^2 \right) +
\sum_{j=1}^{n_1}\sum_{k=n_1+1}^n g \delta(z_{jk}),
\end{eqnarray}
where $\omega$ denotes the angular trapping frequency and
$z_{jk}=z_j-z_k$.
Throughout, we assume $n_1 \ge n_2$.
The solutions for the $(n_1,n_2)=(1,1)$ system
are known semi-analytically for all $g$~\cite{busc98}.
For $n>2$, in contrast,
the eigenenergies and eigenstates are, in general, not known analytically
and we resort to a numerical approach.
To solve the time-independent
Schr\"odinger equation for the Hamiltonian $H$,
we separate the center of mass motion and expand the
Green's function for the relative coordinates
in terms of harmonic oscillator 
states. For the $(2,1)$
system, the approach has been detailed in Ref.~\cite{gharashi}.
For the $(3,1)$ and $(2,2)$ systems, we 
generalize the formalism of 
Refs.~\cite{gharashi,mora,mora-2005,kestner,seesupport}.
Throughout, we assume that the center of mass wave function
is in the ground state and
label the relative eigenstates
by the relative
parity $\Pi^{\rm{rel}}$ ($\Pi^{\rm{rel}}=\pm1$). 
Our calculations 
yield highly-accurate energy spectra and wave functions 
as a function of $g$.
For $g=0$, 
the ground state of
the $(2,1)$ has $\Pi^{\rm{rel}}=-1$, 
that of the $(3,1)$ system has $\Pi^{\rm{rel}}=-1$, and
that of the $(2,2)$ system has $\Pi^{\rm{rel}}=+1$;
in the following,
we restrict ourselves to these subspaces.

\begin{figure}
\vspace*{+.5cm}
\includegraphics[angle=0,width=65mm]{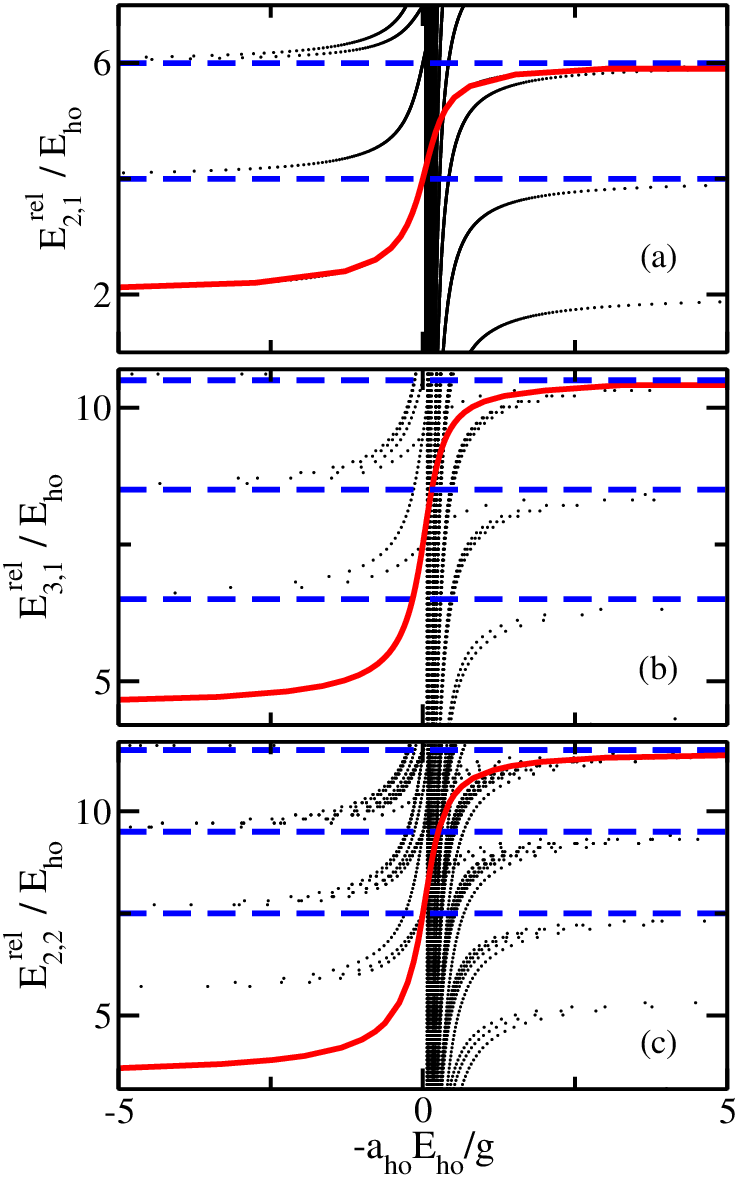}
\vspace*{0.cm}
\caption{(Color online)
Relative energies for the
(a) $(2,1)$ system with $\Pi^{\rm{rel}}=-1$,
(b) $(3,1)$ system with $\Pi^{\rm{rel}}=-1$,
and
(c) $(2,2)$ system with $\Pi^{\rm{rel}}=+1$
as a function of $-1/g$.
The dashed lines show the eigenenergies corresponding to states
that are not affected by the interspecies interactions.
The thick solid lines show the upper branch.}
\label{fig1}
\end{figure}

Figure~\ref{fig1} shows the relative eigenenergies of
the $(2,1)$, $(3,1)$ and $(2,2)$ systems
as a function of $-E_{\rm{ho}} a_{\rm{ho}}/g$,
where $E_{\rm{ho}}$ and $a_{\rm{ho}}$ denote respectively
the harmonic oscillator energy and length,
$E_{\rm{ho}}=\hbar \omega$ and $a_{\rm{ho}}=\sqrt{\hbar/(m \omega)}$.
For $g \rightarrow 0^+$ (far left of the graphs), 
the eigenenergies approach the NI
limit.
As $g$ increases, the eigenenergies increase,
reflecting the repulsive character of the $\delta$-function interactions.
In this work, we are primarily interested in the upper 
branches shown by thick solid lines in Fig.~\ref{fig1}~\cite{seesupport}.
For $1/|g|=0$, the relative energy of the upper branch 
is expected, assuming that some kind of generalized
fermionization takes place, to equal $(n^2-1) E_{\rm{ho}} /2$.
Our numerical energies 
agree with this expectation to better than 0.0001\%,
0.005\% and 0.02\% for the $(2,1)$,
$(3,1)$ and $(2,2)$ systems, respectively~\cite{footnote1}. 
For negative $g$, the spectrum changes notably.
In this regime, the 
upper branch corresponds to a highly excited state
of the model Hamiltonian.
In addition to states whose energies change fairly gradually
with $-1/g$, there exists a set of
``diving states'', reflecting the
fact that the 1d $\delta$-function potential
with negative $g$ supports a two-body bound state. 
The fact that
the two-body binding energy goes to $-\infty$ for $g \rightarrow -\infty$
leads to the accumulation of diving states in Fig.~\ref{fig1} for
small positive $-a_{\rm{ho}}E_{\rm{ho}}/g$.
For positive $g$, the upper branch was mapped out in 
Ref.~\cite{brouzos}.
For negative $g$, the upper branch has been mapped out
for the $(2,1)$ system~\cite{harshman} but not $n>3$.

We now discuss the $(2,1)$ eigenstate
of the upper branch with
$1/|g|=0$.
The energy of the upper
branch of the $(2,1)$ system with $1/|g|=0$ 
is degenerate
with the energy of a state that is not affected by the 
$\delta$-function
interactions [see lowest dashed line in Fig.~\ref{fig1}(a)].
The two degenerate eigenstates
$\psi_{\rm{adia},1}^{|g|=\infty}$ 
and $\psi_{\rm{adia},2}^{|g|=\infty}$ 
[corresponding to the
solid and lowest dashed lines in Fig.~\ref{fig1}(a)]
are, including the center-of-mass contribution, given by~\cite{footnote0}
\begin{eqnarray}
\label{eq_psi21}
\psi_{\rm{adia},1}^{|g|=\infty}=
\frac{a_{\rm{ho}}^{-9/2}}{2 \sqrt{3} \pi^{3/4} }
z_{12}(z_{13} z_{23} - 3 |z_{13}| |z_{23}|) 
f(z_1,z_2,z_3)
\end{eqnarray}
with $f(z_1,\cdots,z_n)=e^{ -\sum_{j=1}^n z_j^2/(2 a_{\rm{ho}}^2) } $
and
$\psi_{\rm{adia},2}^{|g|=\infty}=\psi_{\rm{ideal},0}(z_1,z_2,z_3)$,
where
\begin{eqnarray}
\psi_{\rm{ideal},0}(z_1,z_2,z_3)=\frac{\sqrt{2} a_{\rm{ho}}^{-9/2}}{\sqrt{3} \pi^{3/4}} 
z_{12} z_{13} z_{23} f(z_1,z_2,z_3).
\end{eqnarray}
Since the eigenstate $\psi_{\rm{adia},1}^{|g|=\infty}$ changes 
smoothly when the system Hamiltonian is changed adiabatially
($\psi_{\rm{adia},2}^{|g|=\infty}$ is unchanged),
we
refer to 
these states 
as ``adiabatic eigenstates''.
According to the generalized Fermi-Fermi mapping~\cite{girardeauRapid},
$\psi_{\rm{adia},1}^{|g|=\infty}$ should coincide with 
the state $\psi_{\rm{G},0}$,
which is
obtained by applying the spin-dependent
mapping function
$M_{\rm{FF}}$ ($\sigma_j=\uparrow$ for $j=1,\cdots,n_1$
and $\sigma_j=\downarrow$ for $j=n_1+1,\cdots,n$)~\cite{girardeauRapid},
\begin{eqnarray}
M_{\rm{FF}}=\prod_{1 \le j < k \le n} 
\large[(\delta_{\sigma_j \uparrow} \delta_{\sigma_k \downarrow}-
\delta_{\sigma_k \downarrow} \delta_{\sigma_j \uparrow}) 
\mbox{sgn}(z_{jk})
\nonumber \\
+ \delta_{\sigma_j \uparrow}\delta_{\sigma_k \uparrow}+
\delta_{\sigma_j \downarrow}\delta_{\sigma_k \downarrow}\large],
\end{eqnarray}
to the energetically lowest lying eigenstate
$\psi_{\rm{ideal},0}$ of the trapped NI single-component
Fermi gas.
We find, however, that this is not the case. Instead,
we 
find that $\psi_{\rm{G},0}$ has non-unit overlap with
$\psi_{\rm{adia},1}^{|g|=\infty}$ and $\psi_{\rm{adia},2}^{|g|=\infty}$,
i.e.,
$|\langle \psi_{{\rm{adia}},j}^{|g|=\infty}|\psi_{\rm{G},0} \rangle|^2=8/9$
and $1/9$ for $j=1$ and 2, respectively.

The $(3,1)$ and
$(2,2)$ systems with $1/|g|=0$ support
respectively two and four degenerate states 
with $E^{\rm{rel}}=15 E_{\rm{ho}} /2$.
For the $(3,1)$ system, both states are affected by the
$\delta$-function interactions.
For the $(2,2)$ system, three of the four states are  
affected by the
$\delta$-function interactions.
We find 
$|\langle \psi_{\rm{adia},1}^{|g|=\infty}|\psi_{\rm{G},0} \rangle|^2=
4/5$ 
and $0.865(7)$
for the $(3,1)$ and $(2,2)$ systems, respectively~\cite{footnote0}.
This indicates that
$\psi_{\rm{G},0}$ is, for $n>2$ and $n_1-n_2>0$, a linear combination of
the $\psi_{{\rm{adia}},j}^{|g|=\infty}$ ($j=1,2,\cdots$).
Thus, starting in the energetically lowest lying eigenstate
of the NI system, an adiabatic sweep 
from $g=0^+$ to $g\rightarrow \infty$ does not only lead to 
population of the ``fermionized state'' $\psi_{\rm{G},0}$
but also to population of one or more additional states that are orthogonal
to $\psi_{\rm{G},0}$.

\begin{figure}
\vspace*{+.5cm}
\includegraphics[angle=0,width=65mm]{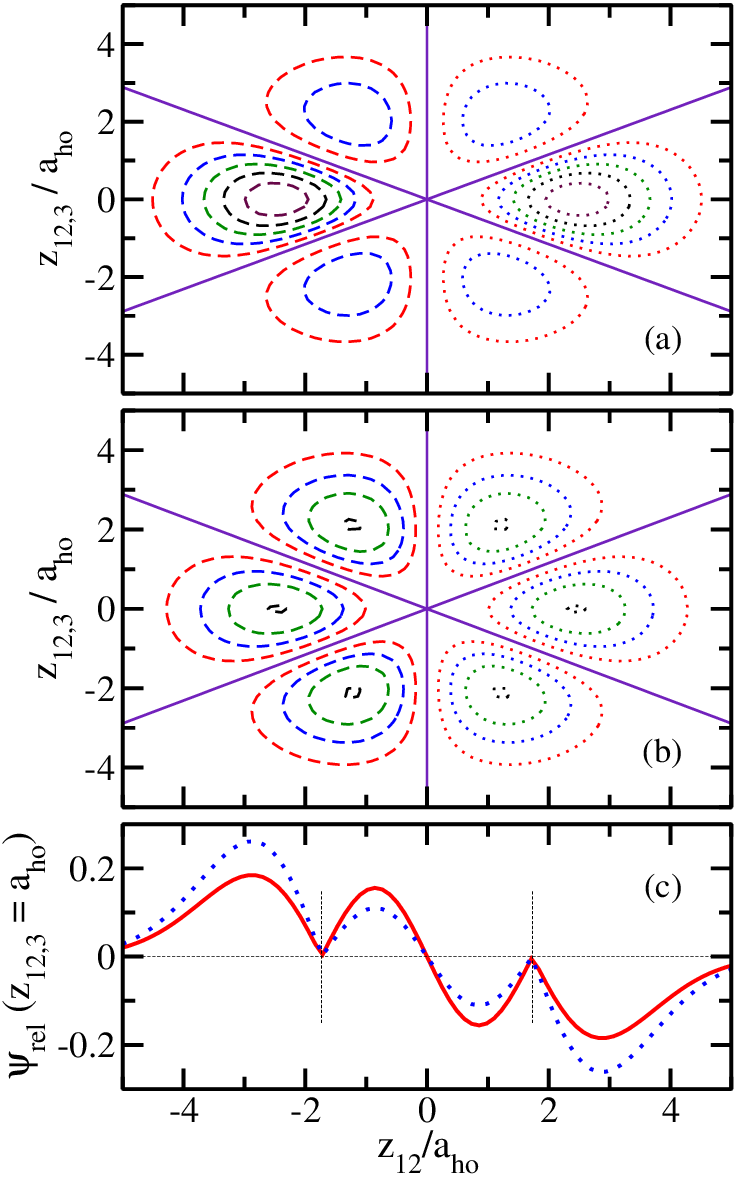}
\vspace*{0.cm}
\caption{(Color online)
Relative wave function of the $(2,1)$ system with 
$1/|g|=0$ and $\Pi^{\rm{rel}}=-1$.
Contour plots of (a) $\psi_{\rm{adia},1}^{|g|=\infty}$
and (b) $\psi_{\rm{G},0}$ as functions of $z_{12}$ and $z_{12,3}$.
Nodal lines are shown by solid lines.
The dashed and dotted contours indicate positive and negative wave 
function regions; the contours are spaced equidistant.
(c) Dotted and solid lines show cuts of 
$\psi_{\rm{adia},1}^{|g|=\infty}$ and $\psi_{\rm{G},0}$
as a function of $z_{12}$ for $z_{12,3}=a_{\rm{ho}}$.
The thin dashed vertical lines at $z_{12}=\pm \sqrt{3}a_{\rm{ho}}$ 
are shown as 
a guide to the eye.}
\label{fig2}
\end{figure}

Figures~\ref{fig2}(a) and \ref{fig2}(b)
show contour plots of the wave functions $\psi_{\rm{adia},1}^{|g|=\infty}$ 
and $\psi_{\rm{G},0}$, respectively, for the $(2,1)$
system with $1/|g|=0$
as functions of the up-up distance coordinate $z_{12}$ and 
the Jacobi coordinate $z_{12,3}$, $z_{12,3}=(z_{13}+z_{23})/ \sqrt{3}$.
The most striking feature is that 
$\psi_{\rm{G},0}$
appears to have a higher ``symmetry'' than 
$\psi_{\rm{adia},1}^{|g|=\infty}$. This is highlighted in the
eigen function cuts shown in Fig.~\ref{fig2}(c).
The absolute value
of the slope of the wave function $\psi_{\rm{G},0}$ near
the nodes at $z_{12}=\pm \sqrt{3}a_{\rm{ho}}$, corresponding to 
$z_{13}=0$ and $z_{23}=0$, is the same 
to the left and right of the node [see solid
line in Fig.~\ref{fig2}(c)].
Mapping $\psi_{\rm{G},0}$ so that it is anti-symmetric
with respect to $z_{13}=0$ 
and $z_{23}=0$ and 
describing the interspecies interactions
through $\delta'$-functions in first-order 
perturbation theory, we find 
$E / E_{\rm{ho}} \approx 9/2+c E_{\rm{ho}} a_{\rm{ho}}/(g\sqrt{2 \pi})$
with $ c=9$.
From our numerical results, in contrast,
we extract 
$c=81/8$.
This discrepancy highlights that the generalized Fermi-Fermi mapping
cannot, in general,
be
utilized within a perturbative
framework.
Figure~\ref{fig2}(c) shows that
the wave function $\psi_{\rm{adia},0}$ is neither symmetric nor
anti-symmetric in the vicinity of 
$z_{13}=0$ and $z_{23}=0$. 
This reflects the fact that the interspecies
degrees of freedom of the two-component Fermi gas 
with 
$n>2$ 
are not constrained by
symmetry. 

\begin{figure}
\vspace*{+.5cm}
\includegraphics[angle=0,width=65mm]{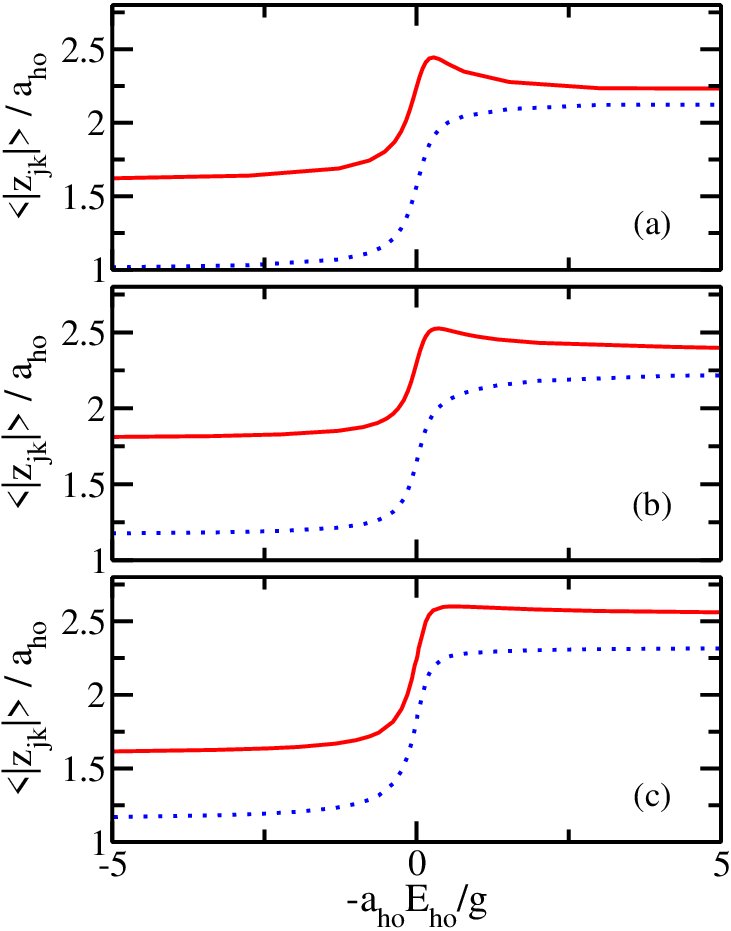}
\vspace*{3.cm}
\caption{(Color online)
Expectation values $\langle |z_{jk} | \rangle$ 
for the up-up (solid lines) and up-down distance 
coordinate (dotted lines) as a function of $-1/g$
for the upper branches of the
(a) $(2,1)$,
(b) $(3,1)$, and
(c) $(2,2)$ systems.
}\label{fig3}
\end{figure}

Next, we discuss the correlations of
the upper branch of the $(2,1)$,
$(3,1)$ and $(2,2)$ systems.
Figure~\ref{fig3} shows the expectation
values $\langle |z_{12}| \rangle $
and  $ \langle |z_{1n}| \rangle$ as a function of $-1/g$.
The expectation value $\langle |z_{1n}| \rangle$
of the up-down distance coordinate 
increases monotonically 
with increasing $-1/g$ for all three systems considered.
The expectation value $\langle |z_{12}| \rangle$
of the up-up distance coordinate, in contrast, first increases
monotonically with increasing $-1/g$, reaches a 
maximum at $g_{\rm{c}}$ ($g_{\rm{c}}<0$)
and then decreases monotonically.
The ``critical'' coupling strengths are 
$-a_{\rm{ho}}E_{\rm{ho}}/g_{\rm{c}} \approx 0.3$, $0.35$ and $0.6$ for the
$(2,1)$, $(3,1)$ and $(2,2)$ systems, respectively.

The energy of the upper branch increases monotonically 
with increasing $-1/g$, suggesting that the {\em{effective}}
interspecies interactions 
for the upper branch are  repulsive 
for all $g$
($g$ positive and negative)
and increase
with increasing $-1/g$.
In a naive picture, this suggests that the system expands with
increasing $-1/g$. Indeed, this is the case for $-1/g \lesssim -1/g_{\rm{c}}$,
as indicated by the fact that $\langle |z_{12}| \rangle$
and $\langle |z_{1n}| \rangle$ increase monotonically in this regime 
with increasing $-1/g$.
However, $\langle |z_{12}| \rangle$ turns around at $g_{\rm{c}}$,
indicating that the system favors smaller distances
between like particles.
For $-1/g \gtrsim -1/g_{\rm{c}}$,
the interspecies 
interactions are so strong that they are more important
than the effective repulsion due to
the Pauli pressure.
An analogous energy competition
drives, according to the Stoner model~\cite{stoner}, 
the transition from a paramagnetic
phase to an itinerant ferromagnetic phase at a critical
interaction strength.
The metastable upper branch has been populated 
experimentally for
small highly-elongated two-component Fermi
gases~\cite{zuernThesis}. These experiments
suggest that decay to lower lying
molecular states is negligibly small even for negative coupling
constants $g$, thereby opening the possibility to study
the correlations discussed above experimentally.

In summary,
we have solved the Schr\"odinger equation
for harmonically confined two-component
Fermi gases in one dimension as a function of the strength
of the interspecies $\delta$-function
interaction. Highly accurate energy spectra
were obtained for the $(2,1)$, $(3,1)$ and $(2,2)$
systems with
positive and negative
interspecies coupling constant.
The strict 1d spectra agree
to about 1\% or better with those
of quasi-1d atomic Fermi gases
with aspect ratio 10 or higher~\cite{gharashi,blume}, which are currently being
investigated by
means of radiofrequency and tunneling spectroscopy
in Jochim's cold atom laboratory in Heidelberg~\cite{zuernPRL,zuernThesis}.
We reported on two characteristics of the upper  
branch: (i) Although the energy of the upper 
branch coincides with that of a fully fermionized system
for infinitely large coupling constant $g$, the corresponding eigenstate
populated by adiabatically changing the system Hamiltonian
does not coincide with that obtained by applying the generalized Fermi-Fermi
mapping proposed in Ref.~\cite{girardeauRapid}.
The underlying rationale is that the states of the upper
branch for
$n>2$ and $n_1-n_2 \ge1$ are more than one-fold degenerate
and that the wave function 
between unlike fermions is not constrained by symmetry 
considerations~\cite{footnote2}.
(ii) 
We calculated the pair-correlations of the upper branch
and found that the expectation value $\langle |z_{12}| \rangle$
associated with the intraspecies distance coordinate exhibits a maximum
for negative $g$. This, combined with the fact that the
expectation value $\langle |z_{1n}| \rangle$
associated with the interspecies distance coordinate increases monotonically
with increasing $-1/g$, indicates an intricate interplay between the 
interspecies interactions and the Pauli exclusion principle,
similar to the energy competition of the Stoner model that
describes the transition from paramagnetic to ferromagnetic behavior.

{\em{Note:}} After submission of this paper for publication,
three 
related manuscripts appeared on the 
arXiv~\cite{zinner,conduit,lewenstein}.

{\bf{Acknowledgments:}}
We acknowledge stimulating discussions and correspondence
with Selim Jochim and his group members.
We also acknowledge
fruitful discussions with Chris H. Greene on how to
connect results obtained 
within the Schr\"odinger equation and mean-field equation frameworks,
and Jason Ho and Liming Guan for pointing out Ref.~\cite{guan} to us.
Support by the ARO
is gratefully acknowledged.
This work was additionally supported by the National Science
Foundation through a grant for the Institute for
Theoretical Atomic, Molecular and Optical Physics
at Harvard University and Smithsonian Astrophysical Observatory.


\begin{thebibliography}{100}
\bibitem{footnote1}
The deviations from the expected values are within our
estimated numerical uncertainties for all three systems considered.
\bibitem{footnote2}
As pointed out in Ref.~\protect\cite{girardeauRapid}, 
the $(n_1,n_2)=(1,1)$ eigenstates 
for $1/|g|=0$ are correctly described by the generalized 
Fermi-Fermi mapping. In fact, in this case, the Bose-Fermi 
mapping and
the generalized Fermi-Fermi mapping yield the same eigenstates.
\end{thebibliography}
\end{document}


\title{Supplemental material for 
``Correlations of the upper
branch of 1d harmonically trapped 
two-component Fermi gases''}

\author{Seyed Ebrahim Gharashi}
\affiliation{Department of Physics and Astronomy,
Washington State University,
  Pullman, Washington 99164-2814, USA}
\author{D. Blume}
\affiliation{Department of Physics and Astronomy,
Washington State University,
  Pullman, Washington 99164-2814, USA}
\affiliation{ITAMP, Harvard-Smithsonian Center for Astrophysics,
60 Garden Street, Cambridge, Massachusetts 02138, USA}


\maketitle

\subsection{Lippmann-Schwinger equation approach}
This section of the supplemental material tabulates the
relative energies $E_{n_1,n_2}^{\rm{rel}}$
of the upper branch of the
strictly 1d
harmonically trapped $(2,1)$,  $(3,1)$ and $(2,2)$ systems.
The notation and units are the same as those introduced in the
main part of the paper.
To solve the 
relative time-independent Schr\"odinger equation,
we reformulate the problem in terms of the Lippmann-Schwinger equation 
and expand the relative Green's function
in terms of harmonic oscillator states 
(see Refs.~\cite{gharashi,mora,mora-2005,kestner}
for details). 
For the three-particle system,
we use the Jacobi coordinates
$z_{12}$ and $z_{12,3}$, where 
$z_{12,3}=(z_1 + z_2 - 2 z_3)/\sqrt{3}$.
For the four-particle systems,
we use a set of Jacobi coordinates based on an H-tree,
$z_1-z_3$, $z_2-z_4$ and $(z_1+z_3-z_2-z_4)/ \sqrt{2}$.
These coordinates allow for a fairly
straightforward implementation of the permutation symmetry 
under exchange of identical particles.
Taking advantage of the fact that the two-particle Green's function
is known analytically~\cite{gharashi,busc98}, 
the solution reduces to solving a matrix equation
of infinite dimension.
In practice, the matrix is truncated at dimension $N_{\rm{max}}$.
The matrix equation is most readily solved
by inputting the energy. The eigenvalues of the matrix equation 
then yield
the corresponding coupling constants $g$.
Tables~I, II and III provide the relative energies 
$E_{n_1,n_2}^{\rm{rel}}$ of the upper branch
of the $(2,1)$,  $(3,1)$ and $(2,2)$ systems 
as a function of $-1/g$.
The reported $-1/g$ values 
are found by analyzing the
energy spectrum as a function of $N_{\rm{max}}$ and by extrapolating to
the $N_{\rm{max}} \rightarrow \infty$ limit.
The largest $N_{\rm{max}}$ value 
considered for the $(2,1)$,  $(3,1)$ and $(2,2)$ systems
is $48$, $51$
and $85$, respectively.

\begin{longtable}{c c}

\caption[Relative energies of the $(2,1)$ system.]{Relative energies of the $(2,1)$ system.} \label{tab_21} \\

\multicolumn{1}{c}{$-a_{\rm{ho}} E_{\rm{ho}}/g$} & \multicolumn{1}{c}{$E_{2,1}^{\rm{rel}}/E_{\rm{ho}}$} \\ 
\endfirsthead

\multicolumn{2}{c}%
{{ \tablename\ \thetable{} -- continued from previous page}} \\
\multicolumn{1}{c}{$-a_{\rm{ho}} E_{\rm{ho}}/g$} & \multicolumn{1}{c}{$E_{2,1}^{\rm{rel}}/E_{\rm{ho}}$}  \\ 
\endhead

\hline 
\multicolumn{2}{|r|}{{Continued on next page}} \\ 
\hline
\endfoot

\hline \hline
\endlastfoot

\hline \hline
$-\infty$&                      $2$
\\
$-48.4533891(2)$&	$2.0123$ 
\\
$-5.1278666(2)$&	$2.1123$ 
\\
$-2.6150182(3)$&	$2.2123$ 
\\
$-1.7094476(3)$&	$2.3123$ 
\\
$-1.2415419(3)$&	$2.4123$ 
\\
$-0.9549028(4)$&	$2.5123$ 
\\
$-0.7606197(4)$&	$2.6123$ 
\\
$-0.6197006(4)$&	$2.7123$ 
\\
$-0.5123436(4)$&	$2.8123$ 
\\
$-0.4274186(4)$&	$2.9123$ 
\\
$-0.358181(2)$&	$3.0123$ 
\\
$-0.300302(2)$&	$3.1123$ 
\\
$-0.250867(2)$&	$3.2123$ 
\\
$-0.207835(2)$&	$3.3123$ 
\\
$-0.1697307(6)$&	$3.4123$ 
\\
$-0.1354493(4)$&	$3.5123$ 
\\
$-0.1041445(4)$&	$3.6123$ 
\\
$-0.0751464(4)$&	$3.7123$ 
\\
$-0.0479103(4)$&	$3.8123$ 
\\
$-0.0219798(4)$&	$3.9123$ 
\\
$0.0030405(4)$&	$4.0123$ 
\\
$0.0275060(3)$&	$4.1123$ 
\\
$0.0517481(3)$&	$4.2123$ 
\\
$0.07609125(6)$&	$4.3123$ 
\\
$0.1008669(2)$&	$4.4123$ 
\\
$0.1264330(3)$&	$4.5123$ 
\\
$0.1531951(4)$&	$4.6123$ 
\\
$0.1816383(4)$&	$4.7123$ 
\\
$0.2123689(4)$&	$4.8123$ 
\\
$0.2461838(4)$&	$4.9123$ 
\\
$0.2841664(4)$&	$5.0123$ 
\\
$0.3278397(3)$&	$5.1123$ 
\\
$0.3796384(4)$&	$5.2123$ 
\\
$0.4430631(4)$&	$5.3123$ 
\\
$0.5242584(4)$&	$5.4123$ 
\\
$0.6342682(4)$&	$5.5123$ 
\\
$0.7954829(4)$&	$5.6123$ 
\\
$1.0613401(3)$&	$5.7123$ 
\\
$1.5990019(3)$&	$5.8123$ 
\\
$3.33765268(4)$&	$5.9123$ 
\\
$\infty$&                       $6$
\end{longtable}

\begin{longtable}{c c}

\caption[Relative energies of the $(3,1)$ system.]{Relative energies of the $(3,1)$ system.} \label{tab_31} \\

\multicolumn{1}{c}{$-a_{\rm{ho}} E_{\rm{ho}}/g$} & \multicolumn{1}{c}{$E_{3,1}^{\rm{rel}}/E_{\rm{ho}}$} \\ 
\endfirsthead

\multicolumn{2}{c}%
{{ \tablename\ \thetable{} -- continued from previous page}} \\
\multicolumn{1}{c}{$-a_{\rm{ho}} E_{\rm{ho}}/g$} & \multicolumn{1}{c}{$E_{3,1}^{\rm{rel}}/E_{\rm{ho}}$}  \\ 
\endhead

\hline 
\multicolumn{2}{|r|}{{Continued on next page}} \\ 
\hline
\endfoot

\hline \hline
\endlastfoot

\hline \hline
$-\infty$&                      $9/2$
\\
$-60.652656(2)$&  	  	$4.5123$ 
\\
$-6.497655(4)$&  		$4.6123$ 
\\
$-3.358608(8)$&  	  	$4.7123$ 
\\
$-2.22879(2)$&  		$4.8123$ 
\\
$-1.64619(2)$&  		$4.9123$ 
\\
$-1.29032(2)$&  		$5.0123$ 
\\
$-1.05008(3)$&  		$5.1123$ 
\\
$-0.87672(3)$&  		$5.2123$ 
\\
$-0.74551(3)$&  		$5.3123$ 
\\
$-0.64258(3)$&  		$5.4123$ 
\\
$-0.55951(3)$&  		$5.5123$ 
\\
$-0.49092(3)$&  		$5.6123$ 
\\
$-0.43320(4)$&  		$5.7123$ 
\\
$-0.38383(4)$&  		$5.8123$ 
\\
$-0.34104(4)$&  		$5.9123$ 
\\
$-0.30345(5)$&  		$6.0123$ 
\\
$-0.27010(5)$&  		$6.1123$ 
\\
$-0.24021(5)$&  		$6.2123$ 
\\
$-0.21317(5)$&  		$6.3123$ 
\\
$-0.18851(5)$&  		$6.4123$ 
\\
$-0.16584(5)$&  		$6.5123$ 
\\
$-0.14484(5)$&  		$6.6123$ 
\\
$-0.12525(6)$&  		$6.7123$ 
\\
$-0.10686(6)$&  		$6.8123$ 
\\
$-0.08946(7)$&  		$6.9123$ 
\\
$-0.07291(6)$&  		$7.0123$ 
\\
$-0.05705(6)$&  		$7.1123$ 
\\
$-0.04177(6)$&  		$7.2123$ 
\\
$-0.02694(8)$&  		$7.3123$ 
\\
$-0.01250(7)$&  		$7.4123$ 
\\
$0.00169(6)$&  		$7.5123$ 
\\
$0.01572(6)$&  		$7.6123$ 
\\
$0.02964(5)$&  		$7.7123$ 
\\
$0.04357(5)$&  		$7.8123$ 
\\
$0.05757(5)$&  		$7.9123$  
\\
$0.07170(8)$&  		$8.0123$ 
\\
$0.08612(4)$&  		$8.1123$ 
\\
$0.10090(8)$&  		$8.2123$ 
\\
$0.11610(8)$&  		$8.3123$ 
\\
$0.13181(2)$&  		$8.4123$ 
\\
$0.14824(6)$&  		$8.5123$  
\\
$0.16547(3)$&  		$8.6123$  
\\
$0.18371(2)$&  		$8.7123$ 
\\
$0.20313(2)$&  		$8.8123$ 
\\
$0.22401(2)$&  		$8.9123$ 
\\
$0.24666(2)$&  		$9.0123$ 
\\
$0.27147(3)$&  		$9.1123$ 
\\
$0.29897(3)$&  		$9.2123$ 
\\
$0.32990(3)$&  		$9.3123$ 
\\
$0.36510(3)$&  		$9.4123$ 
\\
$0.40592(3)$&  		$9.5123$ 
\\
$0.45430(3)$&  		$9.6123$ 
\\
$0.51307(4)$&  		$9.7123$ 
\\
$0.58675(4)$&  		$9.8123$ 
\\
$0.68286(3)$&  		$9.9123$ 
\\
$0.81511(3)$&  		$10.0123$ 
\\
$1.01124(3)$&  		$10.1123$ 
\\
$1.33745(2)$&  		$10.2123$ 
\\
$2.000864(8)$&  		$10.3123$ 
\\
$4.152703(4)$&  		$10.4123$ 
\\
$\infty$&                       $21/2$
\end{longtable}

\begin{longtable}{c c}

\caption[Relative energies of the $(2,2)$ system.]{Relative energies of the $(2,2)$ system.} \label{tab_22} \\

\multicolumn{1}{c}{$-a_{\rm{ho}} E_{\rm{ho}}/g$} & \multicolumn{1}{c}{$E_{2,2}^{\rm{rel}}/E_{\rm{ho}}$} \\ 
\endfirsthead

\multicolumn{2}{c}%
{{ \tablename\ \thetable{} -- continued from previous page}} \\
\multicolumn{1}{c}{$-a_{\rm{ho}} E_{\rm{ho}}/g$} & \multicolumn{1}{c}{$E_{2,2}^{\rm{rel}}/E_{\rm{ho}}$}  \\ 
\endhead

\hline 
\multicolumn{2}{|r|}{{Continued on next page}} \\ 
\hline
\endfoot

\hline \hline
\endlastfoot

\hline \hline
$-\infty$&  	  	$7/2$ 
\\
$-89.000659(3)$		&	$3.5123$
\\
$-9.57445(3)$		&	$3.6123$
\\
$-4.97170(4)$		&	$3.7123$
\\
$-3.31587(5)$		&	$3.8123$
\\
$-2.46271(8)$		&	$3.9123$
\\
$-1.9421(2)$		&	$4.0123$
\\
$-1.5912(2)$		&	$4.1123$
\\
$-1.3384(2)$		&	$4.2123$ 
\\
$-1.1476(3)$		&	$4.3123$ 
\\
$-0.9983(2)$		&	$4.4123$ 
\\
$-0.8781(3)$		&	$4.5123$ 
\\
$-0.7794(3)$		&	$4.6123$ 
\\
$-0.6966(3)$		&	$4.7123$ 
\\
$-0.6263(3)$		&	$4.8123$ 
\\
$-0.5656(4)$		&	$4.9123$ 
\\
$-0.5129(4)$		&	$5.0123$ 
\\
$-0.4663(4)$		&	$5.1123$ 
\\
$-0.4251(4)$		&	$5.2123$ 
\\
$-0.3882(4)$		&	$5.3123$ 
\\
$-0.3548(5)$		&	$5.4123$ 
\\
$-0.3245(4)$		&	$5.5123$ 
\\
$-0.2969(4)$		&	$5.6123$ 
\\
$-0.2717(5)$		&	$5.7123$ 
\\
$-0.2483(4)$		&	$5.8123$ 
\\
$-0.2267(5)$		&	$5.9123$ 
\\
$-0.2066(5)$		&	$6.0123$ 
\\
$-0.1878(6)$		&	$6.1123$ 
\\
$-0.1700(5)$		&	$6.2123$ 
\\
$-0.1530(4)$		&	$6.3123$ 
\\
$-0.1374(4)$		&	$6.4123$ 
\\
$-0.1223(3)$		&	$6.5123$ 
\\
$-0.1081(8)$		&	$6.6123$ 
\\
$-0.0937(6)$		&	$6.7123$ 
\\
$-0.0813(5)$		&	$6.8123$ 
\\
$-0.0685(5)$		&	$6.9123$ 
\\
$-0.056286(5)$		&	$7.0123$ 
\\
$-0.0442(6)$		&	$7.1123$ 
\\
$-0.0326(5)$		&	$7.2123$ 
\\
$-0.0211(6)$		&	$7.3123$ 
\\
$-0.0099(6)$		&	$7.4123$ 
\\
$0.0012(4)$		&	$7.5123$ 
\\
$0.0120(5)$		&	$7.6123$ 
\\
$0.0228(5)$		&	$7.7123$ 
\\
$0.0336(5)$		&	$7.8123$ 
\\
$0.0444(4)$		&	$7.9123$ 
\\
$0.0552(5)$		&	$8.0123$ 
\\
$0.0662(5)$		&	$8.1123$ 
\\
$0.0774(3)$		&	$8.2123$ 
\\
$0.0887(4)$		&	$8.3123$ 
\\
$0.0975(8)$		&	$8.4123$ 
\\
$0.1110(5)$		&	$8.5123$ 
\\
$0.1223(8)$		&	$8.6123$  
\\
$0.1354(3)$		&	$8.7123$ 
\\
$0.1479(3)$		&	$8.8123$ 
\\
$0.1610(3)$		&	$8.9123$ 
\\
$0.1747(2)$		&	$9.0123$ 
\\
$0.1909(8)$		&	$9.1123$
\\
$0.20389(1)$		&	$9.2123$ 
\\
$0.2197(2)$		&	$9.3123$ 
\\
$0.23636(7)$		&	$9.4123$ 
\\
$0.2540(2)$		&	$9.5123$ 
\\
$0.2731(2)$		&	$9.6123$ 
\\
$0.2939(2)$		&	$9.7123$ 
\\
$0.316(2)$		&	$9.8123$ 
\\
$0.3411(2)$		&	$9.9123$
\\
$0.3684(3)$		&	$10.0123$ 
\\
$0.3991(3)$		&	$10.1123$ 
\\
$0.4338(3)$		&	$10.2123$ 
\\
$0.4737(3)$		&	$10.3123$
\\
$0.5202(3)$		&	$10.4123$
\\
$0.5752(3)$		&	$10.5123$
\\
$0.6416(4)$		&	$10.6123$
\\
$0.7240(4)$		&	$10.7123$
\\
$0.8295(4)$		&	$10.8123$
\\
$0.9695(3)$		&	$10.9123$
\\
$1.1653(3)$		&	$11.0123$
\\
$1.4603(3)$		&	$11.1123$
\\
$1.9575(3)$		&	$11.2123$
\\
$2.9796(2)$		&	$11.3123$
\\
$6.15090(2)$		&	$11.4123$
\\
$\infty$&  		$23/2$ 
\end{longtable}

\subsection{$\psi_{\rm{adia}}$
for $1/|g|=\infty$ for the $(2,1)$, $(3,1)$ and $(2,2)$ systems}
This subsection discusses
the construction of the adiabatic eigenfunctions
of the $(2,1)$, $(3,1)$ and $(2,2)$ systems
for infinitely large coupling constant $g$.

We first consider the $(2,1)$ system.
Our calculation yields 
$\psi_{\rm{adia},1}^{|g|=\infty}$ 
in numerical form.
The wave function $\psi_{\rm{adia},2}^{|g|=\infty}$,
which  corresponds to the
state that is not affected by the interactions 
[see dashed line at
$E_{2,1}^{\rm{rel}}=4 E_{\rm{ho}}$
in Fig.~1(a) of the main text], is known
analytically.
This allows us to write
$\psi_{\rm{G},0}= c_{\rm{adia},1} \psi_{\rm{adia},1}^{|g|=\infty} +
c_{\rm{adia},2} \psi_{\rm{adia},2}^{|g|=\infty}$,
where $c_{\rm{adia},2}$ can be found analytically
and
$c_{\rm{adia},1}$ can be obtained numerically
by calculating the overlap between 
$\psi_{\rm{adia},1}^{|g|=\infty}$ (known in numerical form) and
$\psi_{\rm{G},0}$.
Plugging 
$|c_{\rm{adia},2}|^2=1/9$
into $|c_{\rm{adia},1}|^2+|c_{\rm{adia},2}|^2=1$,
we find 
$|c_{\rm{adia},1}|^2=8/9 \approx 0.888888$.
For comparison, our numerically determined 
value is $|c_{\rm{adia},1}|^2=0.888886(2)$.
Lastly, using
$\psi_{\rm{adia},1}^{|g|=\infty} =
\left( \psi_{\rm{G},0}-
c_{\rm{adia},2} \psi_{\rm{adia},2}^{|g|=\infty} \right) /
c_{\rm{adia},1}$, we find the analytical expression
given in Eq.~(2) of the main text.
The analytical wave functions deduced from 
our numerical approach agree with those obtained via a 
group theoretical approach.
Application of Eqs.~(4) and (5) of Ref.~\cite{guan}
yields $\psi_{\rm{adia},1}^{|g|=\infty}$ [in fact,
it yields a linear combination of three
wave functions, where the spin-up and spin-down
particles are ordered
up-up-down (the labeling chosen in the present work),
up-down-up and down-up-up].
In the language of Ref.~\cite{guan},
$\psi_{\rm{adia},2}^{|g|=\infty}$
is associated with a ferromagnetic spin state and given by Eq.~(3)
of Ref.~\cite{guan}.
For completeness, we 
also report the normalized adiabatic $(2,1)$ wave function 
with even parity for $1/|g|=0$,
\begin{eqnarray}
\frac{a_{\text{ho}}^{-9/2}}{2 \pi^{3/4}} \;
z_{12}\bigg( |z_{13}| \; z_{23} +  z_{13} \; |z_{23}| \bigg)
f(z_1,z_2,z_3).
\end{eqnarray}

While the group theoretical approach of Ref.~\cite{guan}
directly yields the adiabatic eigenstates $\psi_{\rm{adia},1}^{|g|=\infty}$
and $\psi_{\rm{adia},2}^{|g|=\infty}$ for the $(2,1)$ systems,
this is {\em{not}} the case for the $(3,1)$ and $(2,2)$ systems.
For the $(3,1)$ system, e.g., application of 
Eqs.~(4) and (5) of Ref.~\cite{guan} yields $\psi_{\text{G},0}$.
To obtain 
the adiabatic eigenstates of the $(3,1)$ and $(2,2)$ systems,
appropriate linear combinations of the states
constructed using the approach of Ref.~\cite{guan}
have to be taken. In the following, we report the adiabatic eigenstates
of the energetically lowest-lying degenerate $(3,1)$ and $(2,2)$ 
energy manifolds for $1/|g|=0$. 

For the $(3,1)$ system, the 
normalized adiabatic eigenstates with odd parity read
\begin{eqnarray}
\psi_{\text{adia},1}^{|g|=\infty} =
\frac{a_{\text{ho}}^{-8}}{3 \pi \sqrt{10} } \; z_{12} \; z_{13} \; z_{23}
\bigg(
|z_{14}| \;  z_{24} \;  z_{34} +
 z_{14} \; |z_{24}| \; z_{34} +
 z_{14} \;  z_{24} \; |z_{34}| -
5 |z_{14}| \; |z_{24}| \; |z_{34}| 
\bigg)
 f(z_1,z_2,z_3,z_4)
\end{eqnarray}
and 
\begin{eqnarray}
\psi_{\text{adia},2}^{|g|=\infty} =
\frac{\sqrt{2} a_{\text{ho}}^{-8}}{3  \pi \sqrt{5} } \; 
z_{12} \; z_{13} \; z_{23}
\bigg(
|z_{14}|  \; z_{24}  \; z_{34} +
 z_{14}  \; |z_{24}| \; z_{34} +
 z_{14} \;  z_{24} \; |z_{34}| 
\bigg)
 f(z_1,z_2,z_3,z_4).
\end{eqnarray}
We find 
$|\langle \psi_{\text{adia},j}^{|g|=\infty} | \psi_{\text{G},0} \rangle |^2 = 
4/5$ and $1/5$ for $j=1$ and $2$, respectively.
For completeness, we also report the adiabatic eigenstates of the
two even parity states,
\begin{eqnarray}
\frac{a_{\text{ho}}^{-8}}{3 \pi \sqrt{2} } 
\; z_{12} \; z_{13} \; z_{23}
\bigg(
 z_{14} \;  z_{24} \;  z_{34} -
|z_{14}| \; |z_{24}| \; z_{34} -
|z_{14}| \;  z_{24} \; |z_{34}| -
 z_{14}  \; |z_{24}| \; |z_{34}| 
\bigg)
 f(z_1,z_2,z_3,z_4)
\end{eqnarray}
and
\begin{eqnarray}
\psi_{\text{ideal},0}= 
\frac{\sqrt{2}a_{\text{ho}}^{-8}}{3 \pi} \; z_{12} \;
z_{13} \; z_{14} \; z_{23} \; z_{24} \; z_{34} 
 f(z_1,z_2,z_3,z_4).
\end{eqnarray}

For the $(2,2)$ system, the 
normalized adiabatic eigenstates with even parity read
\begin{eqnarray}
\psi_{\text{adia},1}^{|g|=\infty} =
\frac{a_{\text{ho}}^{-8}}{12 \sqrt{13} \pi} \; 
z_{12} \; z_{34}  \;
\bigg\{
\left[
 - 8 c_0 -3 \sqrt{3} (1 - c_0^2)^{1/2}
\right]
z_{13} \; z_{14} \; z_{23} \; z_{24}
+ \nonumber \\
\left[
6 c_0 - \sqrt{3} (1 - c_0^2)^{1/2}
\right]
\bigg(
|z_{13}| \; |z_{14}| \; z_{23} \; z_{24} +
|z_{13}| \; z_{14} \; |z_{23}| \; z_{24} + \nonumber \\
|z_{13}| \; z_{14} \; z_{23} \; |z_{24}| +  
z_{13} \; |z_{14}| \; |z_{23}| \; z_{24} + 
z_{13} \; |z_{14}| \; z_{23} \; |z_{24}| +
z_{13} \; z_{14} \; |z_{23}| \; |z_{24}| 
\bigg) \nonumber \\
+ 13 \sqrt{3} (1 - c_0^2)^{1/2}
|z_{13}| \; |z_{14}| \; |z_{23}| \; |z_{24}| 
\bigg\}
 f(z_1,z_2,z_3,z_4),
\label{eq_22-adia1}
\end{eqnarray}
\begin{eqnarray}
\psi_{\text{adia},2}^{|g|=\infty} =
\frac{a_{\text{ho}}^{-8}}{2 \sqrt{6} \pi} \; 
z_{12} \; z_{34}  \;
\bigg(
|z_{13}| \; |z_{14}| \; z_{23} \; z_{24} - 
z_{13} \; |z_{14}| \; z_{23} \; |z_{24}| - \nonumber \\
|z_{13}| \; z_{14} \; |z_{23}| \; z_{24} +
z_{13} \; z_{14} \; |z_{23}| \; |z_{24}| 
\bigg)
f(z_1,z_2,z_3,z_4),
\end{eqnarray}
\begin{eqnarray}
\psi_{\text{adia},3}^{|g|=\infty}= 
\frac{a_{\text{ho}}^{-8}}{12 \sqrt{13} \pi} \; 
z_{12} \; z_{34}  \;
\bigg\{
\left[
3 \sqrt{3} c_0 - 8 (1 - c_0^2)^{1/2}
\right]
z_{13} \; z_{14} \; z_{23} \; z_{24}
+ \nonumber \\
\left[
\sqrt{3} c_0 + 6 (1 - c_0^2)^{1/2}
\right]
\bigg(
|z_{13}| \; |z_{14}| \; z_{23} \; z_{24} +
|z_{13}| \; z_{14} \; |z_{23}| \; z_{24} + 
|z_{13}| \; z_{14} \; z_{23} \; |z_{24}| +  \nonumber \\
z_{13} \; |z_{14}| \; |z_{23}| \; z_{24} + 
z_{13} \; |z_{14}| \; z_{23} \; |z_{24}| +
z_{13} \; z_{14} \; |z_{23}| \; |z_{24}| 
\bigg) \nonumber \\
- 13 \sqrt{3} c_0
|z_{13}| \; |z_{14}| \; |z_{23}| \; |z_{24}| 
\bigg\}
 f(z_1,z_2,z_3,z_4)
\label{eq_22-adia3}
\end{eqnarray}
and
\begin{eqnarray}
\psi_{\text{adia},4}^{|g|=\infty} =
\psi_{\text{ideal},0}.
\end{eqnarray}
Here, 
\begin{eqnarray}
c_0 = \sqrt{1- \frac{\left(3 \sqrt{6} c+2 \sqrt{1-9c^2/8}
\right)^2}{52}}
\end{eqnarray}
with
$c = |\langle \psi_{\text{adia},1}^{|g|=\infty} | \psi_{\text{G},0} \rangle|$
is 
determined numerically.
We find 
$|\langle \psi_{\text{adia},j}^{|g|=\infty} | \psi_{\text{G},0} \rangle |^2 \approx
0.865(7)$, $0$, $0.023(6)$   
and $1/9$ for $j=1,2,3$ and $4$, respectively.
For completeness, we also report the adiabatic eigenstates of the
two odd parity states,
\begin{eqnarray}
\frac { a_{\text{ho}}^{-8}}{\sqrt{30} \pi}
z_{12} \; z_{34}
\bigg( 
|z_{13}| \; z_{14} \; z_{23} \; z_{24} +
z_{13} \; |z_{14}| \; z_{23} \; z_{24} +
z_{13} \; z_{14} \; |z_{23}| \; z_{24} +
z_{13} \; z_{14} \; z_{23} \; |z_{24}|
   \bigg)
 f(z_1,z_2,z_3,z_4)
\end{eqnarray}
and
\begin{eqnarray}
\frac { a_{\text{ho}}^{-8}}{4 \sqrt{30} \pi}
z_{12} \; z_{34}
\bigg(
-3 \;
|z_{13}| \; z_{14} \; z_{23} \; z_{24} 
-3 \; z_{13} \; |z_{14}| \; z_{23} \; z_{24} 
-3 \; z_{13} \; z_{14} \; |z_{23}| \; z_{24} 
-3 \; z_{13} \; z_{14} \; z_{23} \; |z_{24}|  + \nonumber \\
5 \;
z_{13} \; |z_{14}| \; |z_{23}| \; |z_{24}| +
5 \; |z_{13}| \; z_{14} \; |z_{23}| \; |z_{24}| +
5 \; |z_{13}| \; |z_{14}| \; z_{23} \; |z_{24}| +
5 \; |z_{13}| \; |z_{14}| \; |z_{23}| \; z_{24} 
   \bigg)
 f(z_1,z_2,z_3,z_4).
\end{eqnarray}